\documentclass[twocolumn]{aastex63}
\usepackage{CJKutf8}

\begin{document}

\title{The Velocity Dispersion Function for Massive Quiescent and Star-Forming Galaxies at 0.6 $<$ z $\leq$ 1.0}

\author[0000-0003-0705-6691]{Lance Taylor}
\affiliation{Department of Physics and Astronomy and PITT PACC, University of Pittsburgh, Pittsburgh, PA 15260, USA}

\author[0000-0001-5063-8254]{Rachel Bezanson}
\affiliation{Department of Physics and Astronomy and PITT PACC, University of Pittsburgh, Pittsburgh, PA 15260, USA}

\author[0000-0002-5027-0135]{Arjen van der Wel}
\affiliation{Sterrenkundig Observatorium, Universiteit Gent, Krijgslaan 281 S9, 9000 Gent, Belgium}

\author[0000-0001-9820-9619]{Alan Pearl}
\affiliation{Department of Physics and Astronomy and PITT PACC, University of Pittsburgh, Pittsburgh, PA 15260, USA}

\author[0000-0002-5564-9873]{Eric F. Bell}
\affiliation{Department of Astronomy, University of Michigan, 1085 South University Ave., Ann Arbor, MI 48109, USA}

\author[0000-0003-2388-8172]{Francesco D’Eugenio}
\affiliation{Sterrenkundig Observatorium, Universiteit Gent, Krijgslaan 281 S9, 9000 Gent, Belgium}

\author[0000-0002-8871-3026]{Marijn Franx}
\affiliation{Leiden Observatory, Leiden University, P.O.Box 9513, NL-2300 AA Leiden, The Netherlands}

\author[0000-0003-0695-4414]{Michael V. Maseda}
\affiliation{Leiden Observatory, Leiden University, P.O.Box 9513, NL-2300 AA Leiden, The Netherlands}

\author[0000-0002-9330-9108]{Adam Muzzin}
\affiliation{Department of Physics and Astronomy, York University, 4700 Keele St., Toronto, Ontario, M3J 1P3, Canada}

\author[0000-0001-8823-4845]{David Sobral}
\affiliation{Department of Physics, Lancaster University, Lancaster LA1 4YB, UK}

\author[0000-0001-5937-4590]{Caroline Straatman}
\affiliation{Sterrenkundig Observatorium, Universiteit Gent, Krijgslaan 281 S9, 9000 Gent, Belgium}

\author[0000-0001-7160-3632]{Katherine E. Whitaker}
\affiliation{Department of Astronomy, University of Massachusetts, Amherst, MA, 01003 USA}
\affiliation{Cosmic Dawn Center (DAWN), Copenhagen, Denmark}

\author[0000-0002-9665-0440]{Po-Feng Wu \begin{CJK*}{UTF8}{bkai}(吳柏鋒)\end{CJK*}}
\affiliation{National Astronomical Observatory of Japan, Osawa 2-21-1, Mitaka, Tokyo 181-8588, Japan}

\begin{abstract}
    We present the first direct spectroscopic measurement of the stellar velocity dispersion function (VDF) for massive quiescent and star-forming galaxies at $0.6 < z \leq 1.0$.  For this analysis we use individual measurements of stellar velocity dispersion from high-S/N spectra from the public Large Early Galaxy Astrophysics Census (LEGA-C) survey.  We report a remarkable stability of the VDF for both quiescent and star-forming galaxies within this redshift range, though we note the presence of weak evolution in the number densities of star-forming galaxies.  We compare both VDFs with previous direct and inferred measurements at local and intermediate redshifts, with the caveat that previous measurements of the VDF for star-forming galaxies are poorly constrained at all epochs.  We emphasize that this work is the first to directly push to low-stellar velocity dispersion ($\sigma_\star > 100$ km s$^{-1}$) and extend to star-forming galaxies.  We are largely consistent with the high-sigma tail measured from BOSS, and we find that the VDF remains constant from the median redshift of LEGA-C, $z\sim0.8$, to the present day.
\end{abstract}

\section{Introduction}

Stellar velocity dispersion, $\sigma_\star$, measured as the Gaussian width of the line-of-sight velocity distribution of stars within a galaxy, is a fundamental observable property of galaxies.  It is a key parameter in the fundamental plane of elliptical galaxies \citep{djorgovski87} and can be used to probe galaxy evolution \citep[e.g.,][]{shankar09,chae10,bezanson11,bezanson12}.  Stellar velocity dispersion is correlated with many galaxy properties, including galaxy color and star formation rate \citep{franx08,bell12}, as well as central supermassive black hole (SMBH) mass \citep{shankar09} and dark matter halo properties \citep{zahid18}.  The velocity dispersion of a galaxy can also enable a measurement of dynamical mass via the Virial Theorem \citep{taylor10,cappellari13,zahid&geller17}.  In sum, measuring stellar velocity dispersions unlocks access to a wide range of physical properties.

One way to provide a foundation for studying these scaling relations is to measure how galaxies change through cosmic time as a function of velocity dispersion.  The velocity dispersion function (VDF) yields the number density of galaxies as a function of stellar velocity dispersion.  This measurement is important not just for understanding galaxy demographics, but also for a diverse variety of astrophysics such as: interpreting aggregate lensing signals \citep{turner84}, estimating SMBH populations \citep{shankar09}, and constraining cosmological parameters \citep{mitchell05}.  We note that the velocity dispersions used to compute the VDF are typically integrated velocity widths from large fractions of a galaxy, in contrast to dispersions measured within a smaller region which do not tend to include rotation contributions, e.g., those utilized in older fundamental plane works.  Although for local quiescent galaxies this correction might be small, the contribution of rotational support should be more significant for star-forming galaxies \citep[e.g.,][]{vandeSande18} or quiescent galaxies at earlier times \citep{toft17,newman18,bezanson18a}.  The local and low-redshift VDF has been thoroughly measured (local: \citealt{mitchell05,choi07,shankar09,chae10,bezanson11,sohn17}; low-$z$: \citealt{sheth03,bernardi10,bezanson12}), with most studies focusing on early-type/quiescent galaxies.

Measuring individual stellar velocity dispersions is especially challenging as it requires moderate resolution spectra with high signal-to-noise in the stellar continuum (the more easily observable ionized gas lines may often be absent and/or may not necessarily probe the kinematics of the galaxies).  As a result, at intermediate and higher redshifts most works so far have relied on inferring velocity dispersions from photometric data \citep{bezanson11,bezanson12} or have made assumptions about the evolution of the local relation \citep{shankar09,chae10,geng21}, with only \cite{montero-dorta17} spectroscopically measuring the VDF for the most massive red sequence galaxies at $z\simeq 0.55$.  Each respective inference method comes with different, significant systematic uncertainties.  This means that the VDF beyond the local universe is poorly constrained, limiting the use of the evolving VDF in other studies.

The Large Early Galaxy Astrophysics Census (LEGA-C) survey resolves these issues by having a uniform selection method, independent of color or morphology, that results in minimal selection bias.  Moreover, LEGA-C's deep, high-S/N ($\sim$ 20 \AA$^{-1}$) spectra allow for the extraction of stellar kinematics, including robust stellar velocity dispersions for the vast majority of the sample.  Thus, in this paper we are able to present the VDF as directly measured from the spectroscopic stellar kinematics of a uniformly selected sample of quiescent and star-forming galaxies at $0.6 < z \leq 1.0$.

The paper is organized as follows: In Section \ref{sect:sample} we outline the LEGA-C sample and data processing procedures.  Section \ref{sect:vdfs} details our measurements of the VDF for the full sample, in context with the current literature, which showcase the remarkable stability of the VDF for both populations of galaxies.  Finally, in Section \ref{sect:summary} we summarize our findings and discuss their implications.  We assume a flat $\Lambda$CDM cosmology with $H_0 = 70$ km s$^{-1}$ Mpc$^{-1}$, $\Omega_M = 0.3$, and $\Omega_\Lambda = 0.7$.

\section{Data} \label{sect:sample}

This work is based primarily on the LEGA-C survey \citep{vanderWel16,straatman18,vanderWel21} (PI: van der Wel).  LEGA-C is a deep, moderate-resolution spectroscopic survey that targets massive galaxies in the COSMOS field.  Galaxy targets are brighter than $K_{AB} = 20.7 - 7.5$ log$\, (\frac{1 + z}{1.8})$ at $0.6 < z \leq 1.0$, which corresponds to an approximate stellar mass-limit of log$\, (\frac{M_\star}{M_\odot}) \geq 10.4$.  Spectroscopic target selection relies on the \cite{muzzin13a} v4.1 UltraVISTA catalog, and observations produce spectra between $\sim 6300 \ \text{and} \  8800$ \AA \ in the observed frame.  Full descriptions of the survey, data reduction, and quality can be found in \cite{vanderWel16}, \cite{straatman18}, and \cite{vanderWel21}.  Structural parameters, including a best-fitting S\'{e}rsic profile, are derived from the HST/ACS F814W image of each galaxy \citep{koekemoer07,massey10} using \textit{Galfit} \citep{peng02,peng10} following, e.g., \cite{vanderWel12}.  To separate between the quiescent and star-forming populations, we adopt the standard survey categorizations based on rest-frame $U-V$ and $V-J$ colors according to \cite{muzzin13b}.

The stellar velocity dispersions used in this work are derived using the spatially integrated, optimally extracted spectra \citep[see][]{bezanson18a,bezanson18b}, having been fit using \textit{pPXF} \citep{cappellari04,cappellari17} with a non-negative linear combination of theoretical single stellar population templates (C. Conroy, priv. comm) and Gaussian emission lines and broadening.  In this paper, we denote this observed, spatially integrated stellar velocity dispersion as $\sigma_{\star,int}^\prime$ following \cite{bezanson18b}.  For a subset of the LEGA-C dataset, \cite{vanHoudt21} developed Jeans modelling of the spatially resolved stellar kinematics.  That work uses model dynamical masses to derive an average inclination and aperture correction based on projected galaxy shapes.  We refer to this corrected stellar velocity dispersion measurement as $\sigma_{\star,vir}$ (since it can be used alongside semi-major axis length and S\'ersic index to compute the dynamical mass using the Virial Theorem).  We calculate $\sigma_{\star,vir}$ for our full sample as follows: \begin{equation}
\sigma_{\star,vir} = \sigma_{\star,int}^\prime \left(0.87 + 0.39 \cdot e^{-3.78 \left(1 - q\right)}\right)\text{,}
\end{equation} where $q$ is the projected axis ratio.  These corrections are minor, with average values of $\sim0.006$ dex for quiescent galaxies and $\sim0.018$ dex for star-forming galaxies.  $\sigma_{\star,vir}$ is the stellar velocity dispersion we use to measure our VDFs.  We note that $\sigma_{\star,vir}$ does not attempt to disentangle ordered rotation, which would also contribute to the overall spectral broadening and should be accounted for when, e.g., describing the dispersion support of a galaxy.  Some previous studies of the VDF, especially at low-redshift, adopt a definition of stellar velocity dispersion that attempts to disentangle ordered rotation.  Although this measurement is possible for $\sim800$ galaxies in the LEGA-C sample \citep{vanHoudt21}, the decreased sample size and increased measurement uncertainty would yield a less precise measurement of the VDF and would systematically lower number densities at fixed velocity dispersion. Furthermore, the Virial definition of stellar velocity dispersion provides a more complete accounting of the gravitational orbits inside of the galaxies. We emphasize that these definitional choices often lead to significant systematics and likely dominate the offsets amongst VDF studies.

Our sample of 2586 galaxies is created by starting with the primary LEGA-C sample ($\tt\string PRIMARY=1$; 3017 galaxies) from the DR3 catalog and making a redshift cut to only include galaxies within $0.6<z_{spec} \leq 1.0$ (2868 galaxies).  We then remove observations that are not of a single galaxy with regular morphology ($\tt\string FLAG\_MORPH=0$; 2685 galaxies).  The exclusion of non-regular morphology galaxies has negligible effects on our results, with no significant difference between VDFs created with and without this cut.  Because completeness corrections are essential to the current study, we also make a cut to remove galaxies without a total correction value (a reduction of 99 galaxies; see following paragraph for discussion of total correction values).

Careful completeness corrections are enabled by the well-defined selection of the LEGA-C sample (based on K-band magnitudes).  The team derived volume corrections {to transform the primary targets into a volume limited sample.  The maximum redshift at which a galaxy would still be included in the parent sample was calculated and used to derive the standard $V_{max}$ correction.  For sample targeting corrections, because the probability of inclusion in LEGA-C only depends on $K_s-$band magnitude, correction values are merely} the inverse of the fraction of galaxies with a certain $K_s-$band magnitude observed in UltraVISTA.  {Total correction factors, the product of the volume and sample targeting corrections, range from $\sim$ 0.72 to 36, with a median of $\sim$ 2.3} (see \citealt{straatman18} and Appendix A in \citealt{vanderWel21} for {an expanded} discussion of the completeness corrections).

In addition, we introduce a third correction factor to account for incompleteness in stellar velocity dispersion that arises from the K-band magnitude selection of the LEGA-C sample.  This factor corrects for galaxies with $\sigma_{\star} > 100$ km\,s$^{-1}$ that scatter to K-band magnitudes systematically missed by the LEGA-C targeting. To estimate these correction factor, the Faber-Jackson relation derived for the LEGA-C sample \citep{bezanson18b} is assumed to extend to low-stellar masses.  Thus, a mock sample of velocity dispersions can be created using the estimated 0.08 dex scatter in $\sigma_{\star}$ for quiescent galaxies, 0.11 dex scatter for star-forming galaxies, and a scatter uncertainty of 0.01 dex.  We count the number of galaxies fainter and brighter than the K-band magnitude LEGA-C selection limit for a given redshift and $\sigma$ bin, yielding the completeness of the LEGA-C parent sample with respect to the full population in that redshift and $\sigma$ bin.  This process is repeated 3000 times, each time randomly varying the mock stellar velocity dispersions according to the scatter and its uncertainty, providing us with population completeness correction for each galaxy.  These correction factors, which are multiplicatively combined, along with the known effective volume of LEGA-C ($3.664 \times 10^6$ Mpc$^3$), facilitate the translation from galaxy counts to number densities.

\begin{figure*}[!ht]
    \includegraphics[width=\textwidth]{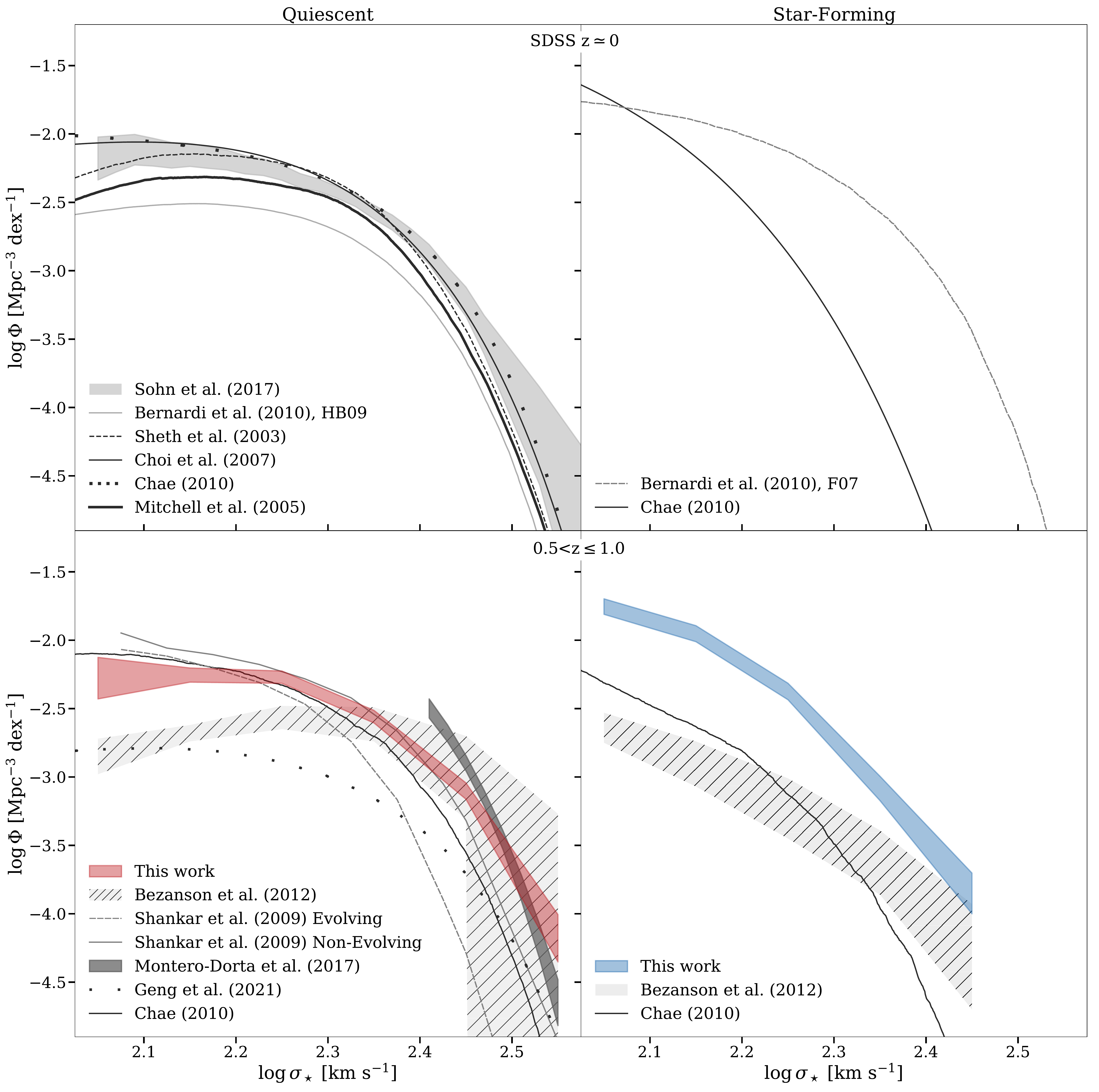}
    \caption{Our measurements of the VDF in the context of previous literature measurements in the local universe (top row) and at higher redshifts (bottom row) and split into quiescent (left) and star-forming (right) populations. Previous measurements vary significantly in sample selection and methodology, including reliance on several indirect measurements of velocity dispersions.  This work (colored bands) represents the first direct measurement of the VDF for the full population of galaxies beyond the local universe, collapsing previous systematic uncertainties in our understanding of the evolving distribution of gravitational potential wells. \label{figs:litcomparison}}
\end{figure*}

\begin{figure*}[!ht]
    \includegraphics[width=\textwidth]{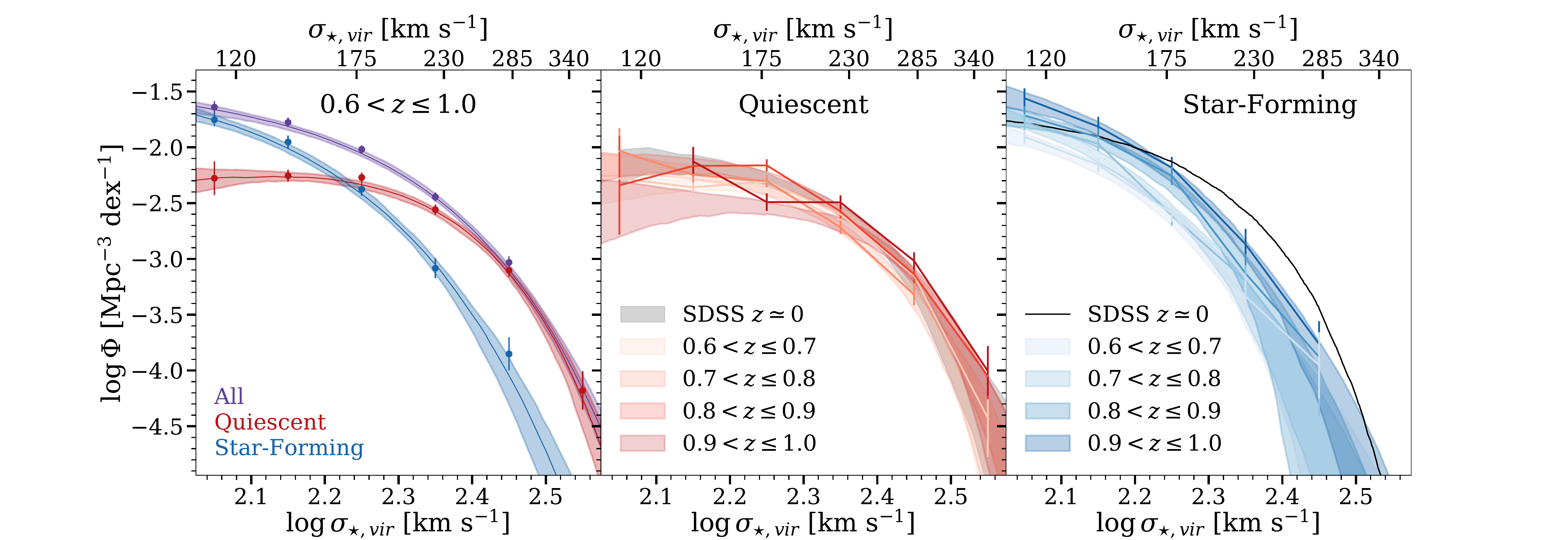}
    \caption{VDFs of quiescent (red), star-forming (blue), and all galaxies (purple) for $0.6 < z \leq 1.0$ (left panel) and in bins of $\Delta z=0.1$ (center and right panels).  The points (left panel) and lines (center and right panels) are the VDFs calculated using the direct spectroscopic measurements from LEGA-C.  The red, blue, and purple bands are $1\sigma$ ranges for the best-fit modified Schechter function.  The grey band (center) and black line (right) are the most recent $z\simeq0$ VDFs from the literature for each respective population (quiescent: \citealt{sohn17}; star-forming: \citealt{bernardi10}).  Both galaxy populations exhibit remarkable stability in $\Phi(\sigma_{\star,vir})$ from $z\sim1$ until today, although there appears to be weak evolution of number densities for star-forming galaxies within the redshift range probed by LEGA-C. \label{figs:vdfs}}
\end{figure*}

\section{Quiescent and Star-Forming Velocity Dispersion Functions} \label{sect:vdfs}

In this section we present the primary results of this study: the number density of galaxies at $z\sim0.8$ as a function of stellar velocity dispersion, or the Velocity Dispersion Function (VDF). We start by showing the current state of the VDF literature in Figure \ref{figs:litcomparison}.  The upper row shows local universe VDFs, which primarily rely on SDSS data. The bottom row shows VDFs calculated at higher redshifts (out to $z\sim1$), including our measurements (colored bands).  Left panels include measurements of the VDF for quiescent/early-type populations, with measurements for star-forming galaxies on the right.

We calculate the number density of quiescent and star-forming galaxies as a function of stellar velocity dispersion as follows, computing $\Phi(\sigma_{\star,vir})$ in bins of 0.1 dex for $\sigma_{\star,vir} \geq 100$ km s$^{-1}$.  Galaxy counts are determined by computing a weighted sum of the relevant sample of galaxies, in which each galaxy is weighted by the total sample completeness corrections {(outlined in Section \ref{sect:sample}}: Vmax, sample targeting, and corrections for scatter between $\sigma_{\star}$ and the K-magnitude LEGA-C survey selection). Number densities and their respective $1\sigma$ uncertainties of our full redshift range quiescent and star-forming $\sigma_{\star,vir}$ VDFs can be found in Table \ref{tbls:vdfs}.
The quoted $1\sigma$ uncertainties on $\Phi$ include Poisson statistics, measurement errors in individual stellar velocity dispersion measurements, and cosmic variance (separately and added in quadrature).  We incorporate $\sigma_\star$ measurement uncertainties via a 1000-iteration bootstrap re-sampling within the formal errors for each galaxy, measuring $\Phi(\sigma_{\star,vir})$ in each iteration, and quote the scatter at each bin. \textit{We note that while cosmic variance is insignificant for most other studies, which probe orders of magnitude larger areas of the sky than the 1.4 deg$^{2}$ COSMOS field, this systematic uncertainty can be significant, or even dominant, in our error budget.} In order to account for cosmic variance, we generated 25 mock light cones using UniverseMachine DR1 data \citep{behroozi19}. These light cones span $0.6 < z < 1.0$ and occupy a solid area of 1.4 square degrees, to match the LEGA-C footprint.  Although mock galaxies in the UniverseMachine model are not assigned stellar velocity dispersions, we rank order by stellar mass and the scatter amongst light cones is then estimated at a given number density from the cumulative and binned stellar mass functions, yielding consistent results in each case, with systematic uncertainties ranging from $0.034$ to $0.15$ [Mpc$^{-3}$ dex$^{-1}$].

Each other work utilizes different criteria to separate galaxy populations and methods of computing velocity dispersions and/or number densities. We note that these sample definitions dominate the overall differences in normalization; e.g., \cite{bernardi10} demonstrated that at $\sigma\sim100$ km\,s$^{-1}$ the inferred number density of early type galaxies can vary by up to an order of magnitude based on the adopted morphological classifications. We summarize the choices made in these studies as follows.  Except for \cite{bezanson12}, \cite{montero-dorta17}, and \cite{sohn17}, all previous work relies on morphological classifications to differentiate between early and late type galaxies.  \cite{bezanson12} uses the same rest-frame $U-V$, $V-J$ color split that is adopted in this paper, and \cite{montero-dorta17} and \cite{sohn17} also split based on slightly different colors to isolate quiescent and star-forming populations.  We note that in addition to complications related to differences in sample definitions, there is a secondary effect introduced by the fact that the definition of stellar velocity dispersion differs amongst studies.  Partially this difference stems from aperture effects, e.g., the difference between $\sigma_{R_e}$ and $\sigma_{R_e/8}$. However, we note that such corrections are minor; adopting the $\sigma(R)$ from \cite{cappellari06}, this corresponds to a net shift of $\sim 0.06$ dex.  \cite{sheth03}, \cite{mitchell05},  \cite{choi07}, \cite{shankar09}, \cite{bernardi10}, \cite{montero-dorta17}, \cite{sohn17}, and \cite{geng21} all use directly measured stellar velocity dispersions (\citealt{shankar09} and \citealt{geng21} use them in the local universe to infer their intermediate redshift results).  On the other hand, \cite{chae10} uses a $L-\sigma$ transformation, while \cite{bezanson12} uses velocity dispersions that are inferred from stellar masses and sizes.  We note that \cite{chae10} and \cite{geng21} use strong-lensing statistics and local SDSS data to construct their intermediate redshift VDFs (though their methods differ), while \cite{shankar09} use age indicators in local SDSS data.  No previous work corrects their velocity dispersions to $\sigma_{\star,vir}$ as is done in this study, and each method is subject to a variety of systematic uncertainties.  However, we emphasize that the differences between each work are subtle and negligible with respect to the broad uncertainties of all prior measurements.  We plot publications that include direct measurements of the VDF and $1\sigma$ error bars in tabular form as bands, and those that report best-fitting modified Schechter functions as lines.

\begin{deluxetable*}{cccccccccccc}[!ht]
\tablecaption{Quiescent and Star-Forming VDFs for $0.6 < z \leq 1.0$. \label{tbls:vdfs}}
\tablecolumns{12}
\tablehead{\colhead{$\log\sigma_{\star,vir}$} & \colhead{$\log\Phi_q$} & \colhead{$\pm(\sigma_{\star,vir})$} & \colhead{$\pm(\text{Poi})$} & \colhead{$\pm(\text{cv})$} & \colhead{$\pm(\text{Tot})$} & \colhead{$\log\Phi_{sf}$} & \colhead{$\pm(\sigma_{\star,vir})$} & \colhead{$\pm(\text{Poi})$} & \colhead{$\pm(\text{cv})$} & \colhead{$\pm(\text{Tot})$} & \colhead{$\log\Phi_{all}$} \\
\colhead{$[\rm{km\,s^{-1}}]$} & \colhead{$[{\rm{Mpc}^{-3}\rm{dex}^{-1}}]$} & & & & & \colhead{$[{\rm{Mpc}^{-3}\rm{dex}^{-1}}]$} & & & & & \colhead{$[{\rm{Mpc}^{-3}\rm{dex}^{-1}}]$}}
\startdata
\vspace{-1.0em} \\
    \multicolumn{12}{c}{$0.6 < z \leq 1.0$} \vspace{0.5em} \\
    \cline{1-12} $2.05$ & $-2.28$ & $0.08$ & $0.03$ & $0.05$ & $0.15$ & $-1.75$ & $0.02$ & $0.01$ & $0.04$ & $0.06$ & $-1.64 \pm 0.06$ \\
    $2.15$ & $-2.25$ & $0.03$ & $0.02$ & $0.04$ & $0.05$ & $-1.95$ & $0.02$ & $0.01$ & $0.04$ & $0.06$ & $-1.78 \pm 0.04$ \\
    $2.25$ & $-2.27$ & $0.02$ & $0.01$ & $0.04$ & $0.05$ & $-2.38$ & $0.03$ & $0.01$ & $0.04$ & $0.06$ & $-2.02 \pm 0.04$ \\
    $2.35$ & $-2.56$ & $0.02$ & $0.01$ & $0.04$ & $0.05$ & $-3.08$ & $0.06$ & $0.03$ & $0.05$ & $0.09$ & $-2.45 \pm 0.04$ \\
    $2.45$ & $-3.10$ & $0.02$ & $0.03$ & $0.05$ & $0.06$ & $-3.85$ & $0.10$ & $0.06$ & $0.08$ & $0.14$ & $-3.03 \pm 0.06$ \\
    $2.55$ & $-4.18$ & $0.09$ & $0.09$ & $0.12$ & $0.17$ & \nodata & \nodata & \nodata & \nodata & \nodata & $-4.18 \pm 0.17$ \\
    \cline{1-12} \vspace{-1.0em} \\
    \multicolumn{12}{c}{$0.6 < z \leq 0.7$} \vspace{0.5em} \\
    \cline{1-12}$2.05$ & $-2.28$ & $0.13$ & $0.04$ & $0.04$ & $0.16$ & $-1.91$ & $0.04$ & $0.02$ & $0.03$ & $0.06$ & $-1.75 \pm 0.07$ \\
    $2.15$ & $-2.36$ & $0.05$ & $0.03$ & $0.04$ & $0.07$ & $-2.16$ & $0.04$ & $0.02$ & $0.04$ & $0.06$ & $-1.95 \pm 0.05$ \\
    $2.25$ & $-2.31$ & $0.03$ & $0.02$ & $0.04$ & $0.05$ & $-2.59$ & $0.06$ & $0.03$ & $0.04$ & $0.08$ & $-2.13 \pm 0.04$ \\
    $2.35$ & $-2.60$ & $0.03$ & $0.03$ & $0.04$ & $0.06$ & $-3.33$ & $0.16$ & $0.07$ & $0.06$ & $0.19$ & $-2.53 \pm 0.06$ \\
    $2.45$ & $-3.15$ & $0.05$ & $0.06$ & $0.05$ & $0.09$ & $-3.95$ & $0.23$ & $0.15$ & $0.09$ & $0.29$ & $-3.09 \pm 0.09$ \\
    $2.55$ & $-4.42$ & $0.20$ & $0.25$ & $0.15$ & $0.36$ & \nodata & \nodata & \nodata & \nodata & \nodata & $-4.42 \pm 0.36$ \\
    \cline{1-12} \vspace{-1.0em} \\
    \multicolumn{12}{c}{$0.7 < z \leq 0.8$} \vspace{0.5em} \\
    \cline{1-12}$2.05$ & $-2.03$ & $0.10$ & $0.05$ & $0.05$ & $0.20$ & $-1.78$ & $0.04$ & $0.02$ & $0.03$ & $0.07$ & $-1.59 \pm 0.09$ \\
    $2.15$ & $-2.25$ & $0.05$ & $0.03$ & $0.04$ & $0.07$ & $-1.97$ & $0.03$ & $0.02$ & $0.03$ & $0.07$ & $-1.78 \pm 0.05$ \\
    $2.25$ & $-2.30$ & $0.03$ & $0.02$ & $0.04$ & $0.06$ & $-2.61$ & $0.08$ & $0.03$ & $0.04$ & $0.10$ & $-2.13 \pm 0.05$ \\
    $2.35$ & $-2.72$ & $0.03$ & $0.03$ & $0.04$ & $0.06$ & $-3.19$ & $0.11$ & $0.06$ & $0.05$ & $0.13$ & $-2.59 \pm 0.06$ \\
    $2.45$ & $-3.32$ & $0.04$ & $0.07$ & $0.05$ & $0.10$ & $-4.05$ & $0.34$ & $0.15$ & $0.11$ & $0.38$ & $-3.25 \pm 0.10$ \\
    $2.55$ & \nodata & \nodata & \nodata & \nodata & \nodata & \nodata & \nodata & \nodata & \nodata & \nodata & \nodata \\
    \cline{1-12} \vspace{-1.0em} \\
    \multicolumn{12}{c}{$0.8 < z \leq 0.9$} \vspace{0.5em} \\
    \cline{1-12}$2.05$ & $-2.34$ & $0.29$ & $0.09$ & $0.07$ & $0.45$ & $-1.71$ & $0.06$ & $0.02$ & $0.04$ & $0.08$ & $-1.62 \pm 0.11$ \\
    $2.15$ & $-2.17$ & $0.06$ & $0.03$ & $0.04$ & $0.09$ & $-1.91$ & $0.04$ & $0.02$ & $0.04$ & $0.08$ & $-1.72 \pm 0.06$ \\
    $2.25$ & $-2.16$ & $0.03$ & $0.02$ & $0.04$ & $0.06$ & $-2.26$ & $0.05$ & $0.02$ & $0.04$ & $0.08$ & $-1.91 \pm 0.05$ \\
    $2.35$ & $-2.57$ & $0.04$ & $0.03$ & $0.04$ & $0.07$ & $-3.13$ & $0.14$ & $0.05$ & $0.05$ & $0.17$ & $-2.47 \pm 0.06$ \\
    $2.45$ & $-3.14$ & $0.04$ & $0.05$ & $0.05$ & $0.08$ & $-3.88$ & $0.17$ & $0.12$ & $0.08$ & $0.22$ & $-3.07 \pm 0.08$ \\
    $2.55$ & $-4.05$ & $0.10$ & $0.14$ & $0.11$ & $0.20$ & \nodata & \nodata & \nodata & \nodata & \nodata & $-4.05 \pm 0.20$ \\
    \cline{1-12} \vspace{-1.0em} \\
    \multicolumn{12}{c}{$0.9 < z \leq 1.0$} \vspace{0.5em} \\
    \cline{1-12}$2.05$ & \nodata & \nodata & \nodata & \nodata & \nodata & $-1.56$ & $0.06$ & $0.03$ & $0.04$ & $0.09$ & $-1.56 \pm 0.09$ \\
    $2.15$ & $-2.13$ & $0.05$ & $0.04$ & $0.05$ & $0.13$ & $-1.81$ & $0.04$ & $0.02$ & $0.04$ & $0.09$ & $-1.64 \pm 0.07$ \\
    $2.25$ & $-2.49$ & $0.06$ & $0.03$ & $0.04$ & $0.08$ & $-2.18$ & $0.04$ & $0.03$ & $0.04$ & $0.10$ & $-2.01 \pm 0.07$ \\
    $2.35$ & $-2.49$ & $0.03$ & $0.03$ & $0.04$ & $0.06$ & $-2.87$ & $0.09$ & $0.05$ & $0.05$ & $0.14$ & $-2.34 \pm 0.06$ \\
    $2.45$ & $-3.02$ & $0.04$ & $0.04$ & $0.05$ & $0.08$ & $-3.76$ & $0.17$ & $0.10$ & $0.08$ & $0.21$ & $-2.95 \pm 0.07$ \\
    $2.55$ & $-4.00$ & $0.15$ & $0.13$ & $0.10$ & $0.22$ & \nodata & \nodata & \nodata & \nodata & \nodata & $-4.00 \pm 0.22$ \\
    \enddata
\tablecomments{The $1\sigma$ uncertainties of the $\log\Phi$ values (Tot) are a combination of a 1000-iteration bootstrap re-sampling within the formal uncertainties of the $\sigma_{\star,vir}$ measurements for each galaxy ($\sigma_{\star,vir}$), Poisson uncertainty (Poi), cosmic variance (cv), and uncertainty in $\sigma_{\star,vir}$ completeness corrections.}
\end{deluxetable*}

\begin{deluxetable*}{lcccc}[!ht]
\tablecolumns{5}
\tablecaption{Best-fit modified-Schechter function parameters for our $\sigma_{\star,vir}$ VDFs, computed using emcee \citep{emcee}. \label{tbl:schechtparams}}
\tablehead{\colhead{Galaxy Type} & \colhead{$\Phi_*$} & \colhead{$\sigma_*$} & \colhead{$\alpha$} & \colhead{$\beta$} \\
& \colhead{$10^{-3}$ [Mpc$^{-3}$]} & \colhead{[km s$^{-1}$]} & &}
\startdata
\vspace{-1.0em} \\
\multicolumn{5}{c}{$0.6 < z \leq 1.0$} \vspace{0.5em} \\
All & $11^{+12}_{-5}$ & $156^{+29}_{-60}$ & $0.8^{+1.4}_{-0.5}$ & $2.19^{+0.45}_{-0.57}$\\
Quiescent & $1.96^{+0.91}_{-0.37}$ & $183^{+50}_{-69}$ & $2.2^{+2.0}_{-1.2}$ & $2.8^{+1.1}_{-0.8}$\\
Star-Forming & $10^{+17}_{-5}$ & $100^{+36}_{-47}$ & $1.0^{+2.0}_{-0.7}$ & $1.80^{+0.60}_{-0.47}$\\
\vspace{-1.0em} \\
\multicolumn{5}{c}{$0.6 < z \leq 0.7$} \vspace{0.5em} \\
Quiescent & $1.8^{+1.9}_{-0.4}$ & $218^{+40}_{-79}$ & $1.7^{+2.3}_{-1.2}$ & $3.7^{+2.0}_{-1.4}$\\
Star-Forming & $7^{+13}_{-4}$ & $108^{+46}_{-50}$ & $0.9^{+1.8}_{-0.6}$ & $2.0^{+1.3}_{-0.6}$\\
\vspace{-1.0em} \\
\multicolumn{5}{c}{$0.7 < z \leq 0.8$} \vspace{0.5em} \\
Quiescent & $2.8^{+3.7}_{-1.1}$ & $172^{+49}_{-80}$ & $1.3^{+2.2}_{-0.9}$ & $2.5^{+1.3}_{-0.9}$\\
Star-Forming & $11^{+27}_{-7}$ & $93^{+42}_{-41}$ & $1.0^{+1.8}_{-0.7}$ & $1.8^{+1.0}_{-0.5}$\\
\vspace{-1.0em} \\
\multicolumn{5}{c}{$0.8 < z \leq 0.9$} \vspace{0.5em} \\
Quiescent & $2.4^{+1.6}_{-0.6}$ & $145^{+64}_{-80}$ & $2.6^{+3.3}_{-1.6}$ & $2.2^{+1.0}_{-0.7}$\\
Star-Forming & $10^{+19}_{-6}$ & $133^{+43}_{-58}$ & $0.8^{+1.9}_{-0.6}$ & $2.5^{+2.1}_{-0.9}$\\
\vspace{-1.0em} \\
\multicolumn{5}{c}{$0.9 < z \leq 1.0$} \vspace{0.5em} \\
Quiescent & $1.6^{+2.3}_{-0.6}$ & $249^{+27}_{-67}$ & $1.1^{+2.2}_{-0.8}$ & $4.1^{+2.3}_{-1.5}$\\
Star-Forming & $21^{+47}_{-13}$ & $115^{+40}_{-50}$ & $0.7^{+1.6}_{-0.5}$ & $2.01^{+0.86}_{-0.61}$\vspace{0.5em}
\enddata
\tablecomments{The values and corresponding $1\sigma$ uncertainty ranges shown are the $16^{\text{th}}$, $50^{\text{th}}$, and $84^{\text{th}}$ percentiles of the samples in the marginalized distributions from the MCMC results.}
\end{deluxetable*}

Looking at the upper left panel of Figure \ref{figs:litcomparison} it can be seen that at high-velocity dispersions (log$\, \sigma_\star\text{/km s$^{-1}$}\simeq2.35$) there is agreement amongst the various local quiescent VDFs of previous works.  All of the prior results in this panel also broadly share a similar VDF shape, although the turnover ($\sigma_*$, see Eq. \ref{eq:schecht}) occurs at slightly different velocity dispersions, creating disagreement at the low-velocity dispersion end.  This is most likely due to differences between how each work categorizes their quiescent/early-type population.  Turning now to the upper right panel of Fig. \ref{figs:litcomparison}, we note that \cite{bernardi10} provide a number of different morphological classifications, and for the sample shown we elect to include Sa, Sb, and Scd galaxies.  Published VDFs of local star forming galaxies exhibit exceptionally large systematic offsets (\citealt{chae10} does not report uncertainties for their modified Schechter function parameters).  This is also the case for the intermediate redshift star-forming VDFs in the bottom right panel.  Both the \cite{chae10} and \cite{bezanson12} star-forming VDFs differ from our low- and medium-dispersion results at the $\sim3\sigma$-level, though we emphasize that these measurements of the VDF are based on inferred velocity dispersions: derived from lensing and masses and sizes, respectively. Given that the classification of star forming galaxies was performed very similarly to the current study \citep{bezanson12}, we expect that this discrepancy is driven by systematically different definitions of $\sigma$. We emphasize that the measurements of $\sigma$ presented in this paper are the more direct, robust quantities.

Moving to the lower left panel, some previous measurements of the intermediate redshift quiescent VDF from extrapolation of local SDSS data \citep{shankar09,chae10} are in tension with our VDF at high-velocity dispersions.  On the other hand, we find good agreement between our quiescent VDF and both the high-mass red sequence VDF from \cite{montero-dorta17}\footnote{We compare only to results that are above the reported $50\%$ completeness level.} and the quiescent VDF from \cite{bezanson12} at high-$\sigma$.  Comparing our results to the VDFs of the first row of Figure \ref{figs:litcomparison}, for quiescent galaxies no significant evolution exists since the median redshift of LEGA-C, $z\sim0.8$, to the present day.  For star-forming galaxies, given lack of consensus in the literature, we cannot draw any conclusions about relative evolution between our intermediate redshift results and those of $z\sim0$.

We turn now to a full discussion of our unique quiescent and star-forming VDF measurements, which are unprecedented beyond the local universe. The left panel of Figure \ref{figs:vdfs} shows our quiescent and star-forming $\sigma_{\star,vir}$ VDFs for the full redshift range of LEGA-C, $0.6 < z \leq 1.0$ (the VDF for all galaxies is shown in purple).  The points with error-bars are the directly measured spectroscopic VDFs using LEGA-C data and the empirical quantities are included in Table 1.  We fit a modified Schechter function \citep{sheth03}: \begin{equation} \label{eq:schecht}
    \Phi(\sigma_\star) \, d\sigma_\star = \Phi_* \, \left(\frac{\sigma_\star}{\sigma_*}\right)^{\alpha} \,  \frac{exp[-(\sigma_\star/\sigma_*)^{\beta}]}{\Gamma(\alpha/\beta)} \, \beta \, \frac{d\sigma_\star}{\sigma_\star},
\end{equation} to each VDF using emcee \citep{emcee}.  These models are shown as lines (in the left panel) and the 16-84\% confidence bands are shown in all panels.  See Table \ref{tbl:schechtparams} for each VDF's best-fit modified Schechter function parameters and the respective $1\sigma$ confidence intervals of each model parameter.  As seen in previous studies at all epochs, star-forming galaxies dominate the population at low-velocity dispersion, while quiescent galaxies are most prevalent at $\sigma_\star\gtrsim 175\, \text{km s}^{-1}$.

In the other two panels of Figure \ref{figs:vdfs} we subdivide the sample into narrower redshift ranges, with quiescent galaxies in the middle panel and star-forming in the right.  In this case, the internal comparison removes the systematic uncertainties in classification and measurement of stellar velocity dispersions that plague comparisons amongst studies in Figure \ref{figs:litcomparison}. Focusing first on the central panel, the most recent quiescent/early-type $z\simeq0$ VDF from the literature, \cite{sohn17}, is shown as a grey band for reference.  {Upon visual inspection} we find no evolution in the number densities of quiescent galaxies, neither within the redshift range probed by LEGA-C nor in comparison with the local VDF.  {We perform two non-parametric tests to validate this conclusion.  First, we compare the number of galaxies above velocity dispersion thresholds, including all targeting and population correction factors.  This test is a model-independent test that is comparable to evaluating evolution in normalization or $\Phi_*$.  Velocity dispersion thresholds are chosen to be the lower bounds of the velocity dispersion bins used to calculate the VDF.  Our second test uses the same velocity dispersion thresholds, but this time investigates whether the shape of the VDF changes with redshift.  The weighted average velocity dispersion value above each threshold is calculated for each redshift range, using the individual corrections as weights.  Then, for each sample of galaxies in a particular redshift range and above a certain velocity threshold, we bootstrap re-sample with replacement (incorporating targeting correction factors) to find the accompanying standard deviation for each mean value.  Both tests yield non-evolution at every velocity dispersion threshold for quiescent galaxies.  We perform similar tests for the star-forming population and report those results in the next paragraph.}

Looking now at the right panel, we again plot a $z\sim0$ VDF for reference (black line) \citep{bernardi10}.  Within the redshift regime probed by LEGA-C we find a hint of decreasing number densities for star-forming galaxies, specifically below $\sigma_{\star,vir} \simeq 150\, \text{km s}^{-1}$, as our VDF amplitude test showed evolution with respect to redshift for the two lowest thresholds (log $\sigma_{\star,vir} = 2.0$ and $2.1$), at the $\sim4.5\sigma$ and $4\sigma$-levels, respectively.  The shape test for star-forming galaxies returned no significant evolution with redshift.  We emphasize that given the variety of potential systematic uncertainties {in this VDF measurement}, we caution drawing any significant conclusions from the tentative trend {that is seen}.

\begin{figure}[!ht]
    \includegraphics[width=0.47\textwidth]{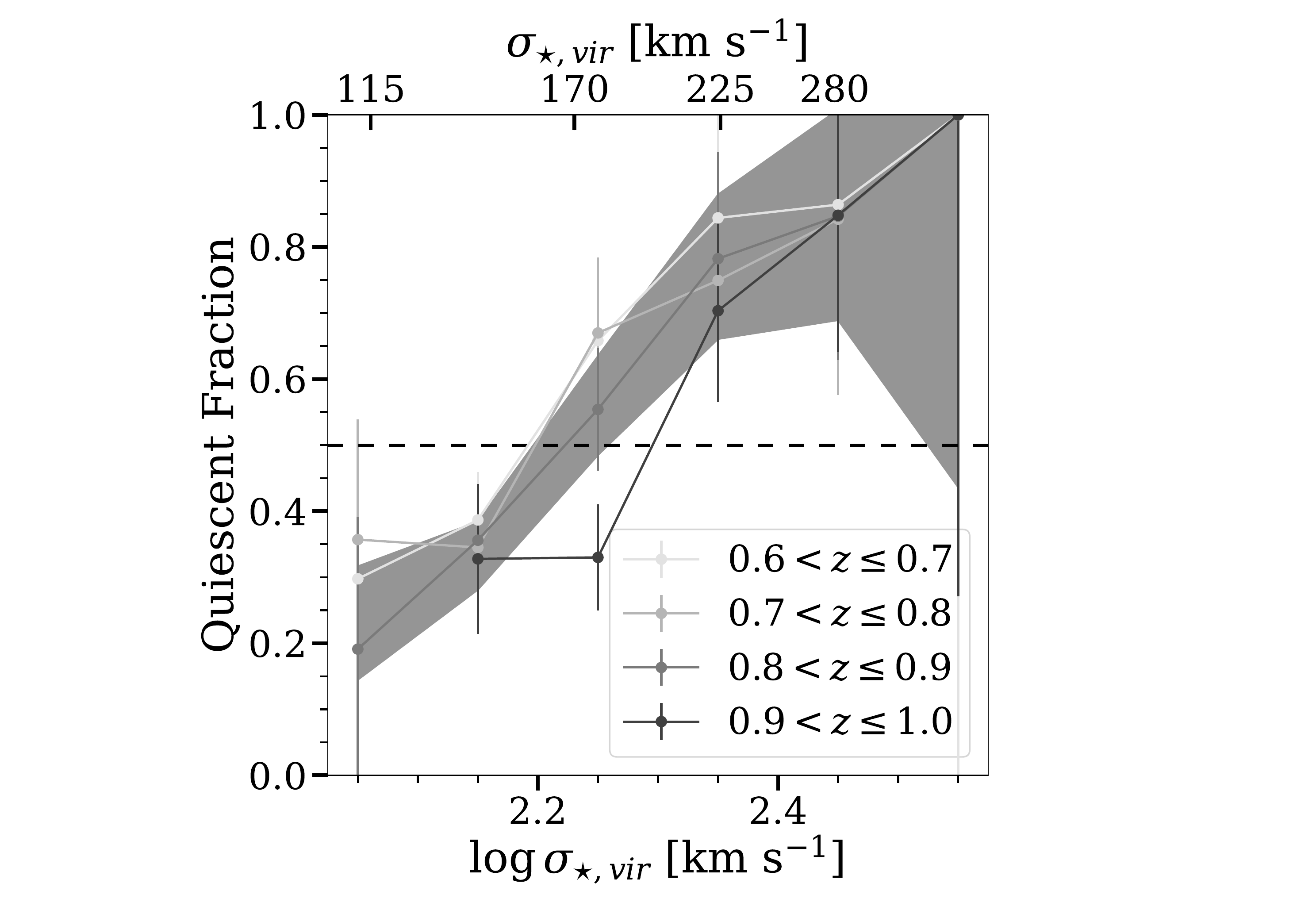}
    \caption{The fraction of quiescent galaxies as a function of log $\sigma_{\star,vir}$.  The quiescent fraction for the full redshift range of LEGA-C ($0.6 < z \leq 1.0$) is shown as a grey band.  Points with error-bars show the function in bins of $\Delta z=0.1$.  The $50\%$ threshold between quiescent or star-forming dominance is designated by the dashed line.  Quiescent galaxies dominate at high-velocity dispersions, while star-forming galaxies are more common at low-sigmas.  The crossover threshold of $\sim$ 170 km s$^{-1}$ may evolve slightly ($\sim$10\%) within the LEGA-C redshift range.}
    \label{figs:qfrac}
\end{figure}

In Figure \ref{figs:qfrac} we explicitly compare the relative fraction of quiescent and star-forming galaxies as a function of velocity dispersion for both LEGA-C's full redshift range, as well as in bins of $\Delta z = 0.1$.  The turnover point between star-forming and quiescent dominated (below and above the dashed line, respectively) is around log $\sigma_{\star,vir} \text{/km s$^{-1}$}\simeq 2.2$ for each relation, except for the highest redshift bin where this value increases to $\simeq 2.3$.

\section{Discussion and Conclusions} \label{sect:summary}

In this paper we demonstrate the remarkable stability of the number densities of both star-forming and quiescent galaxies as a function of directly measured stellar velocity dispersions since $z\sim1$.  This work is made possible by the deep spectra and well-defined targeting strategy of the LEGA-C spectroscopic survey.  We emphasize that above $\sim$100 km s$^{-1}$ the VDF is surprisingly stable.  This result holds within the redshift regime probed directly by LEGA-C ($0.6<z<1.0$) and in comparison with VDFs measured in the local universe.  Furthermore, this stability appears to hold for both the quiescent and star-forming populations of galaxies, although there may be a very weak hint that the number densities of star-forming galaxies decrease with time.  This represents a fundamentally novel measurement of the unchanging distribution of gravitational potential wells beyond the local universe.  {We note that we do not account for possible correlation between the VDF and the environment of the galaxies, a relation that \cite{sohn17g,sohn20} have shown is relevant for our understanding of galaxy formation.}

The stability of the VDF differs from the differential evolution seen in the stellar mass function (SMF), which exhibits weak growth in the population of massive quiescent galaxies and relatively stable number densities of massive star-forming galaxies since $z\sim1$ \citep[e.g.,][]{sobral14,davidzon17}.  This implies that even as quiescent galaxies join the sample, e.g., by growing in mass, they must do so while maintaining a stable central gravitational potential well. This could be explained by inside-out growth via minor mergers \citep{bezanson09,naab09,vanDokkum10}.  That stands in slight contrast to the differential build-up of low-velocity dispersion galaxies presented in \cite{bezanson12}, but we are confident that our current measurements are more robust to systematics than that previous work which relied on inferred velocity dispersions.  Additionally, for star-forming galaxies the hint of subtle number density evolution at fixed velocity dispersion and the non-evolution of the SMF point towards slight growth in size (taking $\sigma \propto \sqrt{\frac{M}{R}}$).

Our results are superficially consistent with the evolution of the simulated halo VDF \citep{weinmann2013}, although we note that those dark matter velocity dispersions were measured at vastly different physical scales from the stellar velocity dispersions of our galaxies.  However, this parallel may suggest a self-similarity between the baryonic and dark matter growth,  which is consistent with the findings of \cite{zahid18} that the velocity dispersion of quiescent galaxies is correlated with both the mass and velocity dispersion of the dark matter halo.  Within that context, our results suggest that there has been very little evolution of the dispersions of the most massive dark matter halos since $z\sim 1$.  On the other hand, the stability of the VDF seen in this work perhaps reflects the imprint of an earlier epoch, during which gravitational potential wells were set.  We note that the evolution of the stellar velocity dispersion function has not been made within cosmological simulations that include baryonic physics, e.g. EAGLE \citep{schaye15} or IllustrisTNG \citep{pillepich18}.  Such a comparison could provide a stringent test of the efficiency of galaxy formation in such models, as velocity dispersion can provide the most model-independent comparisons between simulated and empirical measurements.

Considering the relationship between stellar velocity dispersion and central SMBH mass \citep{shankar09}, our results point toward the most massive SMBHs having been formed early on in the universe and remaining relatively unchanged since at least $z\sim1$ \citep[relatively unchanged within the context of the very steep $M_{\text{BH}} \propto \sigma^5$ relation, e.g.,][]{mcconnell13}.

Pushing the current study to higher redshifts would provide additional information regarding the formation and evolution of massive galaxies.  Advancing past $z\sim1$ will require increased integration times with more powerful infrared-sensitive spectrographs.  We expect surveys using the next generation of massively-multiplexed IR spectrographs on large telescopes, like MOONS \citep{maiolino20} and PFS \citep{takada14}, to provide measurements of the VDF at even earlier times.

\begin{acknowledgments}

We acknowledge the helpful feedback and constructive criticism from the anonymous referee, which greatly improved this paper. LT and RSB would like to thank the Pennsylvania Space Grant Consortium for funding this research and, along with AP, Jeffrey A. Newman, Brett Andrews, Justin Cole, Biprateep Dey, Lina Florez, Yasha Kaushal, Zachary Lewis, David Setton, and Margaret Verrico for meaningful conversations that contributed to this project.  RSB gratefully acknowledges funding for project KA2019-105551 provided by the Robert C. Smith Fund and the Betsy R. Clark Fund of The Pittsburgh Foundation.  FDE acknowledges funding through the H2020 ERC Consolidator Grant 683184.  Based on observations made with ESO Telescopes at the La Silla Paranal Observatory under programme IDs 194-A.2005 and 1100.A-0949 (The LEGA-C Public Spectroscopic Survey).

\end{acknowledgments}

\software{astropy 3.2.2 \citep{astropy13,astropy18},
corner 2.0.1 \citep{corner},
emcee 3.0.0 \citep{emcee},
seaborn 0.10.1 (\url{https://seaborn.pydata.org})}

\end{document}